\documentclass[twocolumn]{aastex6}
\usepackage{graphicx}
\usepackage{placeins}
\usepackage{amsmath}

\newcommand{\kms}{km s$^{-1}$}

\begin{document}

\title{Constraining the Solar Galactic Reflex Velocity Using Gaia Observations of the Sagittarius Stream}

\author{Christian R. Hayes\altaffilmark{1}, David R. Law\altaffilmark{2}, and Steven R. Majewski\altaffilmark{1}}
\altaffiltext{1}{Department of Astronomy, University of Virginia, P.O. Box 400325, Charlottesville, VA 22904-4325}
\altaffiltext{2}{Space Telescope Science Institute, 3700 San Martin Drive, Baltimore, MD 21218}

\email{crh7gs, srm4n@virginia.edu, dlaw@stsci.edu}

\begin{abstract}

Because of its particular orientation around the Galaxy --- i.e., in a plane nearly perpendicular to the Galactic plane and containing both the Sun and Galactic center --- the Sagittarius (Sgr) stream provides a powerful means by which to measure the solar reflex velocity, and thereby infer the velocity of the Local Standard of Rest (LSR), in a way that is independent of assumptions about the solar Galactocentric distance. Moreover, the solar reflex velocity with respect to the stream is projected almost entirely into the proper motion component of Sgr stream stars perpendicular to the Sgr plane, which makes the inferred velocity relatively immune to most Sgr model assumptions.  Using {\it Gaia} DR2 proper motions of $\sim$2,000 stars identified to be Sgr stream candidates in concert with the \citet{lm10} Sgr $N$-body models (which provide a good match to the {\it Gaia} observations) we constrain the solar reflex velocity induced by its orbital motion around the Galaxy to be $\Theta_{\odot} = 253 \pm 6$ km s$^{-1}$. Assuming a solar peculiar motion in the direction of orbital rotation of 12 km s$^{-1}$, and an LSR velocity of 12 km s$^{-1}$ with respect to the local circular speed, the implied circular speed of the Milky Way at the solar circle is $229 \pm 6$ km s$^{-1}$.

\end{abstract}


\section{Introduction}

It has been over thirty years since the International Astronomical Union addressed disparities in derived values of the solar Galactocentric distance, $R_0$, and the Galactic circular rotation velocity at the Sun, $\Theta_0$, of $\pm1$ kpc and $\pm20$ km s$^{-1}$ respectively by recommending the adoption of $R_0 = 8.5$ kpc and $\Theta_0 = 220$ km s$^{-1}$ ``in cases where standardization on a common set of galactic parameters is desirable.''\footnote{The 1985 recommendation of IAU Commission 33, https://www.iau.org/static/resolutions/IAU1985\_French.pdf.}  Despite continued work to establish these parameters, it seems that the dispersions in determinations have not significantly diminished.

Recent estimates of $\Theta_0$ span a wide range, $218-254$ \kms, including those from radio interferometric proper motion measures of star forming regions \citep{bovy09,reid09} and Sgr A* \citep{rb04}, as well as explorations of stellar kinematics from the APOGEE \citep{bovy12} and SEGUE surveys \citep{schonrich12}, or the kinematics of Cepheids \citep{kawata18}; however the latest of these studies have tended toward the middle of that range.

One source of uncertainty is that there is a degeneracy between the inferred $\Theta_0$ and inferred $R_0$, for most measurement methods.  Assessments are further complicated by the fact that the Sun has a peculiar motion with respect to the Local Standard of Rest (LSR), while the LSR may itself be moving with respect to a simple circular orbit around the Galaxy.  Thus assessing $\Theta_0$ is typically entangled with the $\Theta$-directional components of the solar peculiar motion and LSR motion.  The latter are often simplified to the Galactic Cartesian counterparts $V_{\odot,pec}$ and $V_{LSR,pec}$, respectively (since these velocities are often measured using an ensemble of nearby stars), so that, in cylindrical coordinates, the net revolutionary component of the Sun's motion is $\Theta_{\odot} = \Theta_0 + V_{LSR,pec} + V_{\odot,pec}$. The most recently cited values of $V_{\odot,pec}$ hover around 12 km s$^{-1}$, and \citet{bovy12} claim that $V_{LSR,pec}$ may also be as high as 12 \kms.

In this paper we focus on a new measurement of $\Theta_{\odot}$. \citet{maj06} demonstrated that the Sgr stream provides an effective means to measure $\Theta_{\odot}$ (from which $\Theta_{\rm LSR} = \Theta_0 + V_{LSR,pec}$ and $\Theta_0$ can be inferred), independent of the assumed $R_0$ or the precise shape of the Galactic mass distribution, and that avoids having to observe sources in the heavily crowded and dust-extinguished Galactic Center.  The method exploits the favorable orientation of the Sgr stream, which is in a nearly polar orbit that intersects the Galactic plane virtually along the line between the Sun and the Galactic Center.  In this orientation, the Sgr plane provides a non-rotating reference against which the solar motion may be measured, with almost all of the reflex motion imprinted on the proper motions of Sgr stars perpendicular to the stream.  At the time of \citet{maj06}, the available proper motions for known Sgr stream stars were not good enough to apply the method rigorously.  However, exploiting proper motions of $\sim$1-2 mas yr$^{-1}$ accuracy for $\sim$1-2 dozen spectroscopically-confirmed Sgr stream stars in four fields along the Sgr trailing arm, \citet{carlin12} found that their proper motions were best reproduced by models similar to the \citet[][``LM10'' hereafter]{lm10} Sgr destruction models but utilizing an LSR velocity of $264 \pm 23$ \kms. 

The second data release from the ESA-{\it Gaia} mission \citep[{\it Gaia} DR2;][]{gaiadr2} now provides the opportunity make this measurement with a significantly larger sample of stars having proper motions an order of magnitude more precise.

\section{Finding Sagittarius Stream Stars in {\it Gaia}}
\label{obs.sec}

Red giants are ideal tracers of the Sgr stream, because they can readily be seen to large distances. Thus we initially selected stars from the 2MASS catalogue \citep{2mass} with $J$, $H$, and $K_s$ magnitudes between 10 and 13.5, covering a range of colors and magnitudes where giant stars trace the Sgr stream \citep[e.g.,][]{maj03}.  Such stars were selected in a series of rectangular regions (selected to cover Sgr longitudes $|B_{\odot}| \lesssim 20^{\circ}$) on the sky around the trailing arm of the Sgr stream, which collectively span right ascensions from $\alpha = 320^{\circ}$ to $75^{\circ}$ and declinations $-40^{\circ}<\delta<+40^{\circ}$, which corresponds to about 90$^{\circ}$ of the Sgr stream trailing arm.  The selected stars were then cross-matched with the {\it Gaia} DR2 source catalog \citep{gaiadr2} using the CDS X-match service\footnote{\url{http://cdsxmatch.u-strasbg.fr/xmatch}} adopting a 1\arcsec \ positional tolerance.

While {\it Gaia} DR2 does not measure parallaxes to a precision sufficient to reach the distances of the Sgr stream, we can still identify and remove nearby Milky Way (MW) contamination using {\it Gaia} DR2 parallaxes.  Stars with relative parallax uncertainties $\sigma_{\pi}/\pi \leq 0.2$ can provide relatively reliable distance measurements, and almost all such stars in our sample have 1/$\pi < 10$ kpc, considerably closer than the trailing arm of the Sgr stream \citep[which ranges from about 20$-$40 kpc at different stream longitudes;][]{maj03,kop12}.  We therefore remove all of these stars from our sample, and, although it seems counter-intuitive, we keep the stars with poor parallaxes, $\sigma_{\pi}/\pi > 0.2$, which are primarily distant stars that we refer to as our ``distant {\it Gaia} sample''.  To reduce the MW contamination in this distant {\it Gaia} sample further, we also removed stars at $\alpha$ $>$ 50$^{\circ}$ and within 25$^{\circ}$ of the Galactic plane because these sky regions lie near the MW disk where contamination washes out the signature of the Sgr trailing arm.

\begin{figure*}
  \centering
  \includegraphics[scale=0.4,trim = 1.25in 0.75in 1.25in 1.25in, clip]{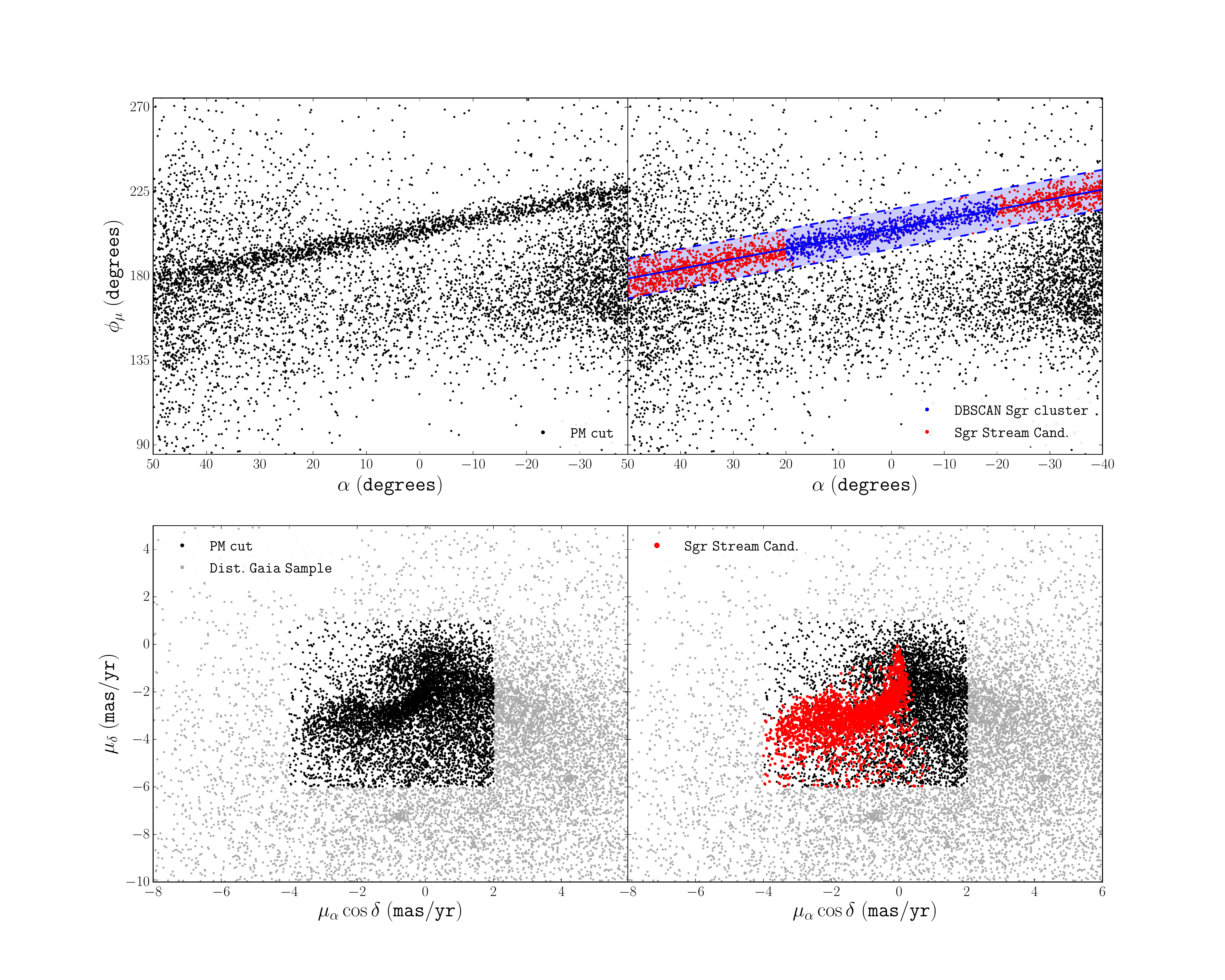}
  \caption{(Bottom) Proper motion vector point diagrams showing (bottom left) the distribution of proper motions in our distant {\it Gaia} sample (grey points), the restricted proper motion selection we use (black points), and the final adopted Sgr trailing arm candidates (red points, bottom right). (Top) Proper motion position angle ($\phi_{\mu}$) as a function of right ascension ($\alpha$), demonstrating (top left) the distinct, linear, and narrow distribution of Sgr trailing arm stars, and (top right) the criteria for selecting our final sample of Sgr trailing arm candidates.   The overplotted solid blue line (top right) is the linear fit to the DBSCAN selected ``Sgr cluster'' (blue points), and the shaded blue region bounded by dashed blue lines is the 3$\sigma$ dispersion around this fit.  The selected Sgr trailing arm candidates include the stars within the latter region (blue points) and those within 3$\sigma$ of the extrapolation of this fit to the plot boundaries (red points).}
  \label{selection}
\end{figure*}

The signature of the Sgr trailing arm is apparent as an arcing overdensity in the {\it Gaia} DR2 proper motion vector point diagram (PMVPD) for our distant {\it Gaia} sample (Fig.~\ref{selection}, lower left panel), so we trimmed in the PMVPD around the overdensity using $-4$ mas yr$^{-1}$ $< \mu_{\alpha}\cos{\delta} < 2$ mas yr$^{-1}$ and $-6$ mas yr$^{-1}$ $< \mu_{\delta} < 1$ mas yr$^{-1}$.  When restricting to these low proper motions, the stars in the Sgr trailing arm stand out with proper motion position angles ($\phi_{\mu}$) that are coherent and nearly linear with right ascension (Fig.~\ref{selection}, top left panel).  The small scatter around this linear trend in $\phi_{\mu}$ arises because the Sgr stream is kinematically cold, and provides an opportunity to refine greatly our selection of Sgr trailing arm candidates. However, elevated MW contamination at some $\alpha$ makes a simple linear fit to the data challenging, even with $\sigma$-clipping to reduce the noise. To avoid this problem, we apply the DBSCAN clustering algorithm \citep[a density-based clustering algorithm that builds clusters of a given density;][]{dbscan} to the the proper motion cut stars between $-20^{\circ} < \alpha < 20^{\circ}$, where the Sgr trailing arm feature is most free of MW contamination.  This allows us to select the Sgr trailing arm feature in a reproducible manner, and use it to extract the feature at right ascensions where the contamination is more considerable.

Setting DBSCAN input parameters $\epsilon = 5$ and $N = 100$, with no normalization of the $\phi_{\mu}$ and $\alpha$ dimensions the DBSCAN algorithm appears clearly to identify a cluster associated with the Sgr trailing arm feature (blue points in the top right panel of Fig.~\ref{selection}).  We then fit a line to this DBSCAN-identified cluster and calculated the dispersion ($\sigma$) in the residuals to this fit. We extend this Sgr stream candidate selection outside the $-20^{\circ}<\alpha<20^{\circ}$ range by extrapolating the fitted line and using the calculated dispersion to select stars within 3$\sigma$ of it.  To measure of the remnant contamination, in the same parameter space we count the number of stars in a region of the same shape/size just above and below our Sgr stream selection, which we average and compare to the number of selected stream candidates to estimate a contamination of about 16\%. This selection therefore includes the DBSCAN selected stars (blue) and the extended sample (red) shown in the right panels of Figure \ref{selection}.

Finally, we trim the selected stars to Sgr orbital longitudes $\Lambda_{\odot} = 30 - 115^{\circ}$ (so that the selection box cuts at right angles across the stream) and reject stars with proper motion errors in $\mu_{\alpha}\cos{\delta}$ or $\mu_{\delta}$ greater than 0.2 mas yr$^{-1}$. This results in a final sample of 1,963 candidate Sgr Stream stars, with an estimated contamination of about 16\%.

The bottom right panel of Figure \ref{selection} shows that the arcing overdensity seen by eye in the PMVPD is, as expected, the feature that corresponds to the Sgr trailing arm candidates selected by their proper motion position angle.  However, as seen in the top panels of Figure \ref{selection}, there is still likely a higher level of contamination at either end of the $\alpha$-range of our sample, particularly at $\alpha \gtrsim 30^{\circ}$.  Moreover, in the $\mu_{\delta}$ distribution of our Sgr trailing arm candidates there appears to be a tail toward values below $\sim -4$ mas yr$^{-1}$.  This low density, low $\mu_{\delta}$ tail is not found in the Sgr stream models discussed below and is likely remnant MW contamination.

\section{N-body Models of the Sagittarius Stream}
\label{nbody.sec}

If the Sgr stream were on a perfectly polar orbit in the Galactic $X-Z$ plane\footnote{We adopt the usual convention in which $XYZ$ describes a left-handed Galactocentric Cartesian coordinate system in which $X$ is positive in the direction of the Galactic anticenter, $Y$ is positive in the direction of Galactic disk rotation at the location of the Sun, and $Z$ is positive towards the North Galactic Pole.} with no orbital precession along the stream, any systematic motion of the stream out of its orbital plane would be entirely due to the Sun's own reflex motion, which could then be measured directly from the data. In reality, Sgr is not on a purely polar orbit, the orbital plane is not precisely aligned with the Galactic $X-Z$ axis, and the observed tidal streams precess slightly with increasing distance from the Sgr core.  Therefore, we must compare the observations against $N$-body models that properly incorporate these various effects.

We do so using the LM10 model, which is well fit to the observed run of Sgr stream angular positions, radial velocities, and distances throughout the $\Lambda_{\odot} = 30 - 115^{\circ}$ range of the trailing arm. Despite the inability of this model to reproduce some key features of the broader Sgr$-$Milky Way system (in particular, the bifurcation of the leading and trailing streams and the large apocentric distance for the trailing arm at $\Lambda_{\odot} \sim 180^{\circ}$) it nonetheless remains the best-constrained model at the orbital longitudes considered here. Indeed, this particular section of the stream is easy to fit {\it regardless} of the assumptions made about the depth and detailed shape of the Galactic gravitational potential \citep[in contrast to the leading arm for which a triaxial halo is necessary to fit both the angular positions and radial velocities simultaneously;][]{ljm09}.  As discussed by LM10 \citep[see also discussion by][]{ljm05}, the dynamically young trailing arm has little power to constrain the shape of the Galactic dark matter halo and can be equally well fit in an oblate, spherical, prolate, or triaxial potential, and regardless of the exact distance to the Sgr core, the distance to the Galactic center, or the overall normalization of the depth of the Galactic gravitational potential via the local circular speed $\Theta_0$.

This insensitivity to these various factors is exactly what makes the trailing arm ideal for this study: For these stars a single and interesting model parameter, $\Theta_0$, is strongly and dominantly coupled to a single observational parameter, one dimension of their proper motion.  In previous studies, for widely varying values of $\Theta_0$, the tangential velocity of the Sgr core along its orbit --- hitherto largely unconstrained given the uncertainties of past proper motion studies --- could be dialed to optimize the match of the implied radial velocities along the trailing stream to available observations.  Now, with the new proper motion constraints available from Gaia, we can discriminate between models having different values of $\Theta_0$ and constrain the solar reflex velocity.

\citet{carlin12} reproduced the LM10 analysis (which assumed $\Theta_0 = 220$ \kms), but for a range of different values of $\Theta_0 = 190 - 310$ \kms\ (sampled every 30 \kms, along with two `best-fit' cases where $\Theta_0 = 232$ \kms\ and 264 \kms).  Because the dark matter halo contributes minimally within the solar circle, these models were realized by scaling the Galactic bulge/disk mass jointly to produce the desired local circular speed.  The details of these models have been discussed previously by LM10 and \citet{carlin12}, and in brief, assume a three-component Galactic mass distribution consisting of a Hernquist spheroid, Miyamoto-Nagai disk, and a triaxial logarithmic dark matter halo with minor/major axis ratio $(c/a)_{\Phi} = 0.72$, intermediate/major axis ratio $(b/a)_{\Phi} = 0.99$, and the minor axis pointing towards $(l,b) = (7^{\circ}, 0^{\circ})$. The Sgr dwarf within this Galactic gravitational potential is constrained to lie at $(l,b) = (5.6^{\circ}, -14.2^{\circ})$, a distance of 28 kpc, and a heliocentric radial velocity $v_{\rm hel} = 142.1$ \kms, with leading and trailing arms that match observed trends of angular position and radial velocity along orbital longitude of the stream.  Each model assumed that the solar peculiar motion with respect to the local circular speed was given by $(U,V,W) = (-9,12,7)$ \kms\ in a left-handed coordinate frame \citep{cox00}, for a range in solar reflex velocities of $\Theta_{\odot} = 202 - 322$ \kms.

\section{Discussion}

\begin{figure*}
  \centering
  \includegraphics[scale=0.6,trim = 0in 0in 0in 0in, clip]{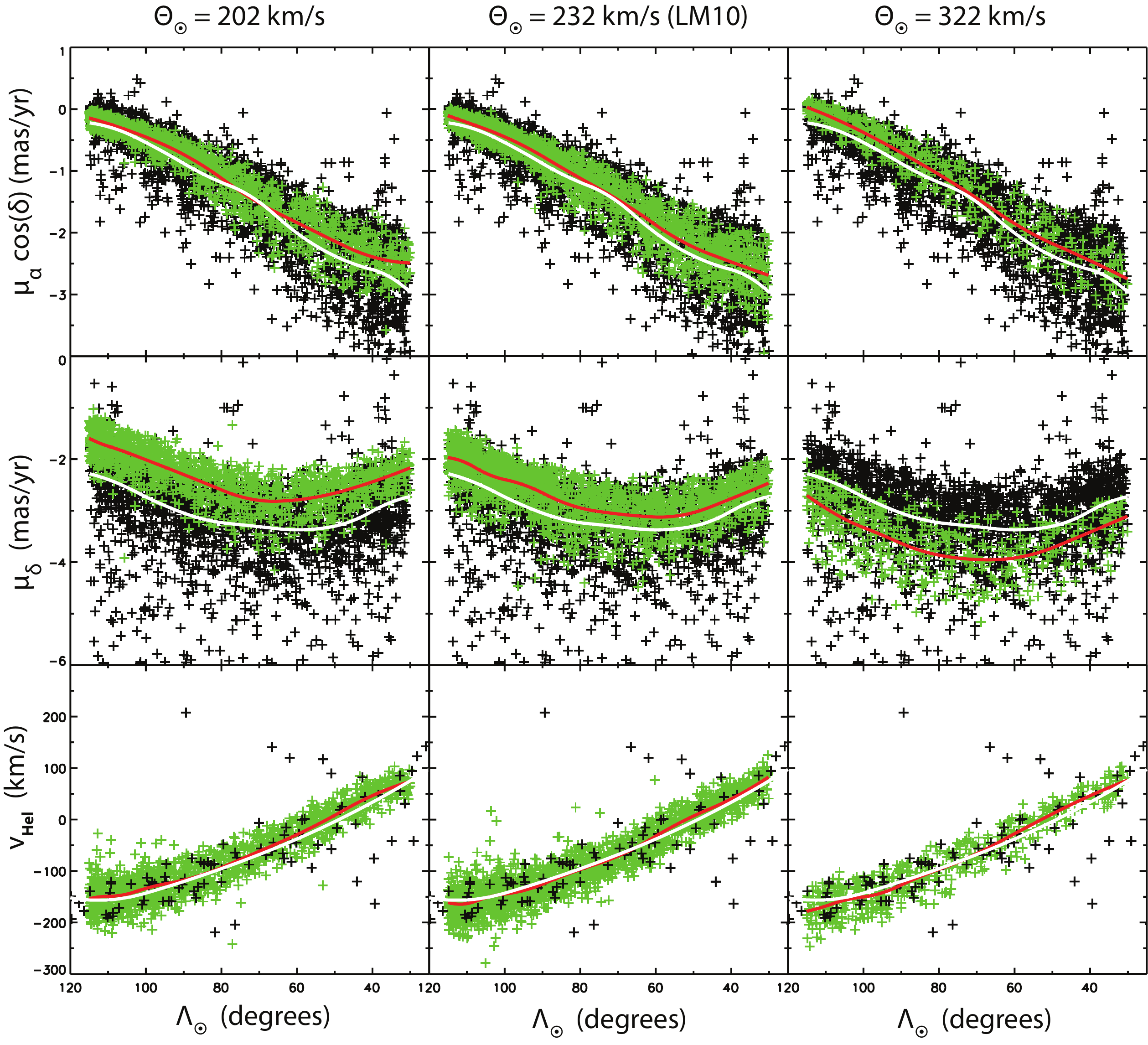}
  \caption{Sgr stream models with different solar reflex velocities compared to observational data as a function of stream longitude $\Lambda_{\odot}$.  The middle column shows the LM10 model, while the left-/right-hand columns show models with slower/faster reflex velocities respectively.  Top row: proper motion along $\alpha$.  Middle row: proper motion along $\delta$.  Bottom row: heliocentric radial velocity.  In the top two rows black points represent the {\it Gaia} observations, while in the bottom row black points represent M-giant observations from \citet{maj04b}.  Green points represent N-body simulated tidal debris.  Solid white/red lines in all panels represent $2\sigma$-clipped spline model fits to the observed/simulated data respectively to guide the eye.}
  \label{nbody.fig}
\end{figure*}

As illustrated in Figure \ref{nbody.fig}, the LM10 model (middle column) continues to be a good match to observations of the Sgr stream in the $\Lambda_{\odot} = 30 - 115^{\circ}$ range of the trailing tail, despite the fact that no proper motions were used to constrain the model.  To a degree this is unsurprising given that LM10 carefully tuned the other four dimensions of phase space to reproduce all observational data available at the time.  We note, however, that the run of proper motions in the declination direction is not quite a perfect match to the {\it Gaia} observations; furthermore, versions of the LM10 model in potentials with slower or faster circular speeds (left and right columns respectively) result in linear shifts of the model in $\mu_{\delta}$ while remaining almost unchanged in $\mu_{\alpha}\cos{\delta}$ and heliocentric radial velocity.

This situation arises because of an additionally fortuitous orientation of the Sgr stream with respect to {\it celestial} coordinates. In $N$-body models where the Milky Way circular speed is larger, the proper motion of model stars in the direction perpendicular to the Sgr plane is larger due to the greater solar reflex motion.  Since the model Sgr dwarf must be made to move faster along its orbit to compensate for the deeper gravitational potential, these model stars also have faster motion within the Sgr plane. In the trailing stream, the vector addition of these components is such that the net change in proper motion for stream stars between different models happens to be almost entirely along $\mu_{\delta}$.

\begin{figure*}
  \centering
  \includegraphics[scale=0.47,trim = 0in 0in 0in 0in, clip]{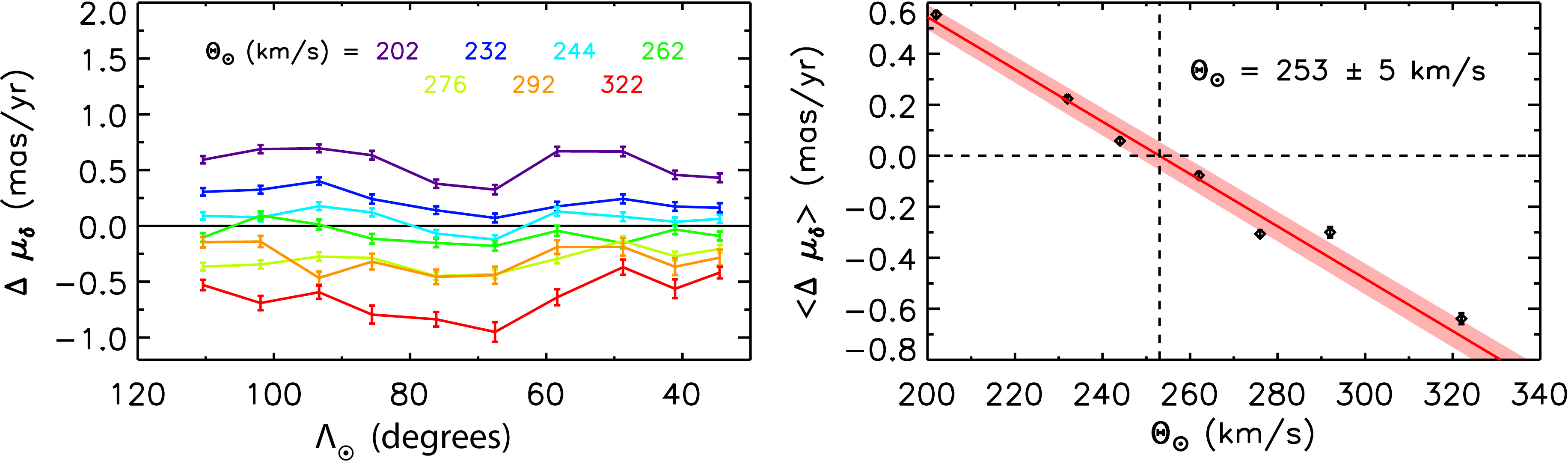}
  \caption{(Left) Difference in $\mu_{\delta}$ proper motion between the observations and the seven $N$-body models (colored lines) in ten bins (each containing the same number of stars) along orbital longitude $\Lambda_{\odot}$.  Error bars represent the 1$\sigma$ uncertainty in the mean for each bin.  (Right) Mean difference in $\mu_{\delta}$ averaged over all longitude bins as a function of solar reflex velocity $\Theta_{\odot}$ (black points with 1$\sigma$ error bars).  The solid red line and shaded red region represent the best first order polynomial fit and associated 1$\sigma$ uncertainty.  The zero-crossing point is located at $\Theta_{\odot} = 253 \pm 5$ \kms.}
  \label{relation.fig}
\end{figure*}

As we show in Figure \ref{relation.fig} (left-hand panel), the difference between the ($2\sigma$-clipped) mean proper motion of the $N$-body models versus the observed stream is nearly constant with $\Lambda_{\odot}$, such that the model with reflex velocity $\Theta_{\odot} = 202$ \kms\ is systematically offset by about +0.5 mas yr$^{-1}$ (purple curve), wheras the model with $\Theta_{\odot} = 322$ \kms\ is offset by about -0.6 mas yr$^{-1}$ (red curve). The consistency of these systematic differences suggests that we can average these offsets over orbital longitude to obtain a single value $\langle \Delta \mu_{\delta} \rangle$ describing the $N$-body model stream offset from the {\it Gaia} proper motions for different $\Theta_{\odot}$ (Fig.~\ref{relation.fig}, right-hand panel).  This relation is well described by a simple linear fit to within observational uncertainty.  By taking this fit (and the $1\sigma$ uncertainties thereon) we solve for the $\langle \Delta \mu_{\delta} \rangle = 0$ \kms\ crossing point and determine that this occurs at $\Theta_{\odot} = 253 \pm 5$ \kms.

An estimate of the possible systematic uncertainty in this measurement can be obtained by comparing the proper motion of the Sgr core in these $N$-body models with the {\it Gaia} observations.  Because the models adopted an orbital pole defined by the path of the tidal streams \citep{maj03}, varying $\Theta_{\odot}$ in these models describes a linear relation in the PMVPD for the Sgr core \citep[see Figure 2.8 of][]{lm16} according to the speed of Sgr along its orbit perpendicular to the line of sight (dialed up and down to compensate for the altered Galactic potential following from changes in $\Theta_0$).  The observed {\it Gaia} proper motion of the Sgr core \citep[$(\mu_{\alpha}\cos{\delta}, \mu_{\delta}) = (-2.692, -1.359)$ mas/yr;][]{helmi18} lies slightly off this relation by about 0.15 mas yr$^{-1}$, but is most consistent with a choice of $\Theta_{\odot} = 256$ \kms.  We therefore adopt 3 \kms\ (the difference between this value and the value derived from fitting the trailing stream) as our systematic uncertainty; by combining systematic and random uncertainty terms our final estimate of the solar reflex velocity is $\Theta_{\odot} = 253 \pm 6$ \kms.

This measurement is consistent with both the values of $\Theta_{\odot} = 242^{+10}_{-3}$ \kms\ obtained by \citet{bovy12}\footnote{$V_{\phi,\odot}$ in their notation.} and $\Theta_{\odot} = 256 \pm 17$ \kms\ obtained by \citet[][using all observable constraints applied to their highest-purity fields]{carlin12} to within $1\sigma$.  Meanwhile, combining the proper motion of Sgr A* in the Galactic Plane, $\mu_{l} = 6.379 \pm 0.026$ mas yr$^{-1}$ \citep{rb04}, with the recent, high precision measure of $R_0 = 8.122 \pm 0.031$ kpc from \citet{gravity} yields a value of $\Theta_{\odot} = 245.6 \pm 1.4$ \kms.  Combining the uncertainties on this $R_0$ dependent measure of $\Theta_{\odot}$ and the $\Theta_{\odot}$ reported here, there is a $1.2\sigma$ difference between the two results, which is still a reasonable agreement. While the reflex motion measured with respect to Sgr A* nominally provides higher precision, our estimate is an important, {\it independent} probe using a method that does not depend on the Galactocentric radius of the Sun.  If we assume that $V_{\odot,pec} = 12$ \kms\ and likewise follow \citet{bovy12} in assuming that $V_{LSR,pec} = 12$ \kms, then our results imply a local circular velocity of $\Theta_0 = 229 \pm 6$ \kms\ for the Milky Way.

We note that the overall 6 \kms\ uncertainty in our estimate of the solar reflex velocity is driven primarily by the large intrinsic width of the stream. In both the observations and $N$-body models the 1$\sigma$ width of the $\mu_{\delta}$ distribution is about 0.5 mas yr$^{-1}$; with ten longitude bins each containing about 200 stars, this translates to an uncertainty of about 11 $\mu$as/yr in the mean.  Similar efforts using dynamically colder streams may therefore be able to obtain more precise results, and such generalizations of our method to arbitrary streams appear promising \citep{Malhan17}.

\acknowledgements
This research made use of the {\sc topcat} \citep{topcat} data visualization tool.

\end{document}